\documentclass[pre,preprint,showpacs]{revtex4-1}
\usepackage{amsfonts}
\usepackage{amssymb}
\usepackage{amsmath}
\usepackage{graphicx}

\begin{document}
\title{ Power-law random banded matrices and ultrametric matrices: eigenvector distribution in the intermediate regime}
\author{E. Bogomolny}
\affiliation{CNRS, Universit\'e Paris-Sud, UMR 8626,
Laboratoire de Physique Th\'eorique et Mod\`eles Statistiques, 91405 Orsay,
France}
\author{M. Sieber}
\affiliation{School of  Mathematics, University of Bristol, University Walk, Bristol BS8 1TW, UK}
\begin{abstract}
The power-law random banded matrices and the ultrametric random matrices are investigated numerically in the regime where eigenstates are extended but all integer matrix moments remain finite in the limit of large matrix dimensions. Though in this case standard analytical tools are inapplicable,  we found that in all considered cases eigenvector distributions are extremely well described by the generalised hyperbolic distribution which differs considerably from the usual Porter-Thomas distribution but shares with it certain universal properties. 
\end{abstract}
\maketitle
%=============================
\section{Introduction}

Random matrix theory (RMT) has an incredibly large field of applications ranging  from nuclear physics to number theory \cite{handbook}. Besides numerous other results of RMT the following two are real benchmarks of the theory: the level repulsion at small distances  and the Porter-Thomas  distribution (PTD) of eigenvectors (see e.g. \cite{porter}, \cite{mehta}). With physical rigour the latter states that  eigenfunctions, $\Psi$, of chaotic quantum systems can be statistically approximated by  Gaussian random variables with zero mean and a variance determined by the normalisation (for simplicity we consider real eigenfunctions)
\begin{equation}
P(\Psi)=\frac{1}{\sqrt{2\pi}}\mathrm{e}^{-\frac{\Psi^2}{2}},\qquad  P(\Psi^2=x)=\frac{\mathrm{e}^{-x/2}}{\sqrt{2\pi x}},\qquad \langle \Psi^2\rangle=1\ .
\label{pt}
\end{equation}
Originally the PTD had been introduced  before the development of RMT  for the description of neutron  resonance widths \cite{porter_thomas}. In the usual invariant random matrix ensembles \cite{mehta}  the PTD \eqref{pt} is a simple consequence of the rotational invariance (plus large matrix dimensions). Recently it has been rigorously proved (see e.g. \cite{knowles}, \cite{tao} and references therein) that for a wide class of non-invariant matrix ensembles  the distribution  of eigenvectors remains the same. 

The simplicity and the universality of the PTD led to 
wide-spread utilisation of this distribution in many different physical contexts.  Surprisingly, some recent experimental results are in contradiction with
Eq.~\eqref{pt} \cite{koehler_1}-\cite{koehler_3}. The authors of these references fitted experimentally measured high resolution data  of $s$-wave neutron widths  to $\chi^2$-distributions with $\nu$ degrees of freedom. They found that the best fit  corresponds to $\nu\approx 0.5-0.6$. As the PTD \eqref{pt} is equivalent  to the $\chi^2$-distribution with $\nu=1$, they estimated from the collection of all their data  that  the probability that the PTD is valid is of the order of $10^{-5}$. Though different scenarios had been proposed to explain such difference  (see e.g. \cite{hans_1}-\cite{bogomolny} among others) it seems that a consensus has not yet been found \cite{hans_2}.

One purpose of this paper is to demonstrate that within RMT there exist models which give rise to eigenfunction distributions different from the PTD but which share similar universal properties.
%==========================
\section{'Physical' random matrix ensembles}

Our starting point  is the (evident) observation that  the usual RMT imposes a non-physical condition that all states interact with each other with approximately the same strength. The most general Wigner matrices for which it is possible to prove universal results are the so-called comparable matrices where each matrix element, $H_{ij}$, is an independent (up to the Hermitian symmetry) Gaussian  random variable with zero mean but with the variance lying between 2 small numbers
\begin{equation}
\frac{C_1}{N}\leq \big \langle H_{ij}^2\big \rangle \leq \frac{C_2}{N},\qquad i,j=1\ldots N
\end{equation}  
where $C_{1,2}$ are constants and $N$ is the matrix dimension. These inequalities are crucial for all applications of standard methods (see e.g. \cite{knowles}, \cite{tao}). 

From physical considerations, models with a certain state hierarchy are much more natural. A typical example of such ensembles  is  the power-law random banded matrices (PLBM) introduced in \cite{mirlin}. Each matrix element of this ensemble is an independent (up to the Hermitian symmetry) Gaussian random variable $H_{ij}$ with zero mean and the variance given by the expressions
\begin{equation}
\langle H_{ii}^2\rangle=2,\qquad \langle H_{ij}^2\rangle_{i\neq j}=a^2(|i-j|)
\label{plbm}
\end{equation}
where the function $a(r)$, $r=|i-j|$,  decreases at large argument as a certain power of the distance from the diagonal
\begin{equation}
a(r)\underset{r\to \infty}{\longrightarrow}\epsilon \, r^{-s}\ .
\end{equation} 
To avoid boundary effects  we choose in calculations an (arbitrary) translation-invariant function
\begin{equation}
a(r)=\epsilon \left [1+\Big (\frac{N}{\pi}\sin( \frac{\pi r}{N} )\Big )^2 \right ]^{-s/2},  \qquad a(r)\underset{r\ll N}{\longrightarrow}\frac{\epsilon}{(1+r^2)^{s/2}}\ .
\label{invariant}
\end{equation}
Other definitions, e.g. $a(r)=\epsilon (1 + (r \,{\textrm{mod}\,  N})^2 )^{-s/2}$, lead to similar results.

\begin{figure}
\begin{center}
\includegraphics[width=.6\linewidth]{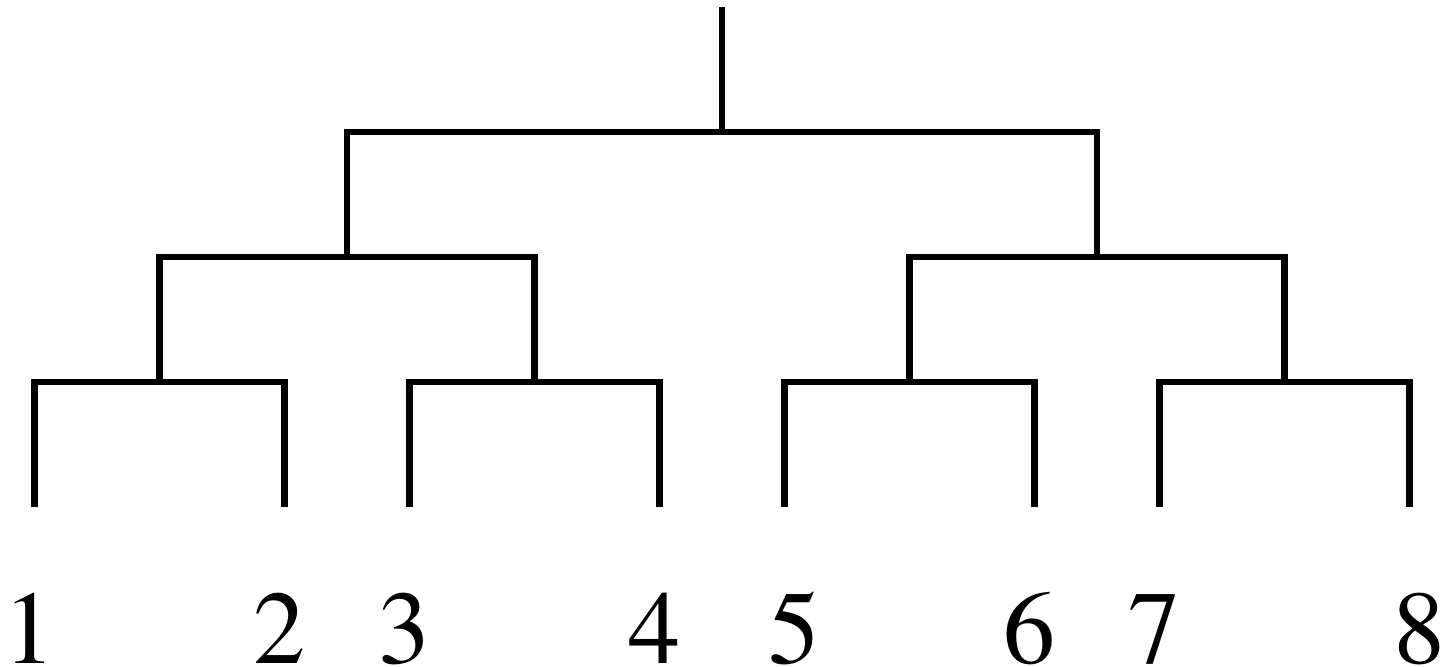}
\end{center}
\caption{The binary tree which determines the ultrametric distance. One has: $\mathrm{dist}(1,2)=1$, $\mathrm{dist}(1,3)=\mathrm{dist}(1,4)=2$, $\mathrm{dist}(1,5)=\mathrm{dist}(1,6)=\mathrm{dist}(1,7)=\mathrm{dist}(1,8)=3$.}
\label{binary_tree}
\end{figure}

The second ensemble which we consider  is the hierarchical analog of PLBM  with the (binary) ultrametric distance between states proposed in \cite{ossipov}. Such ultrametric  matrices (UMM) consists of $2^n\times 2^n$ matrices as above but with the function $a(|i-j|)$ in \eqref{plbm} replaced by
\begin{equation}
a(i,j)=\epsilon \, 2^{-s \, \mathrm{dist}(i,j)}
\label{ultra}
\end{equation} 
where $\mathrm{dist}(i,j)$ is the ultrametric distance  defined as half the number of edges for the shortest path between $i$ and $j$ on a binary tree as in Fig.~\ref{binary_tree}. 

The variance of the diagonal matrix elements is arbitrary. It has been chosen equal to 2 to obtain the  usual GOE matrices in  the case $s=0$ and $\epsilon=1$. 

In both models the parameter $s$ which determines the decrease of the interaction with the distance plays a predominant role.
Define two moments of the matrix $H_{ij}$ by
\begin{equation}
S_1(N)=\frac{1}{N}\sum_{i,j=1}^N \langle \big | H_{ij} \big |\rangle ,\qquad S_{2}(N)=\frac{1}{N}\sum_{i,j=1}^N\langle \big | H_{ij} \big |^2\rangle .
\end{equation}
The rule of thump for such models is the following. If $\lim_{N\to\infty} S_1(N)$  is finite, all eigenvectors are localised and the spectral statistics is Poissonian. If $\lim_{N\to\infty} S_2(N)$ diverges, the eigenvectors are fully delocalised and the spectral statistics is GOE.  For the above models it means that for $s>1$ states are localised, and for 
$s< \frac{1}{2} $ the models (after rescaling) are equivalent to the GOE  ensemble.  In \cite{mirlin} these results   were proved (with physical rigour) for PLBM. In \cite{ossipov} they were only mentioned and the main attention was given to the critical case $s=1$. Recently \cite{warzel}, these statements for the UMM \eqref{ultra} have been rigorously established. 

These simple answers leave the region where $S_1(N)$ diverges but $S_2(N)$ converges unexplored. For PLBM and UMM it corresponds to the following interval of $s$
\begin{equation}
\frac{1}{2}<s<1 .
\label{interval}
\end{equation}
In \cite{mirlin} it  was conjectured  that  in this interval all states in PLBM  are extended, but it appears that no detailed investigation has been performed so far.  

The investigation of such a regime is important also from another point of view. Recently it was argued \cite{kravtsov} that in certain models there may exist within the delocalised phase a new non-ergodic phase characterised by non-trivial fractal dimensions.  The interval \eqref{interval} is the only  place where such non-ergodic states are possible for PLBM and UMM.

%====================================
\section{Results}

The main difficulty in studying the intermediate case \eqref{interval} is related to the absence of a small or large parameter which is required in all standard analytical approaches to random matrix studies. For example, it is clear that mean values of all integer moments of the initial matrix 
\begin{equation}
 \mu_n=\frac{1}{N} \big \langle \mathrm{Tr}\, H^n\big \rangle
\end{equation}
are finite in the limit $N\to\infty$ provided that the condition $s>\frac{1}{2}$ is fulfilled. As no $N$ dependence has been introduced in the matrix variances (cf. \eqref{plbm} and \eqref{ultra}) it implies  that diagonal and off-diagonal terms give comparable contributions though the latter are much more  numerous. It should be compared to the usual  cases where the Wigner semicircle law is a consequence of the dominance of off-diagonal terms. Consequently even the mean eigenvalue density is not known analytically when condition \eqref{interval} is valid. In general properties of convergent series with random coefficients may and will be quite different from divergent series  where the central limit theorem can be applied (cf. e.g. the Bernoulli convolution \cite{bernoulli}). 

In the absence of analytical methods we performed numerical calculations of eigenvector distributions for both, the PLBM and the UMM. In particular, we calculated eigenvalues, $E_{\alpha}$, and corresponding (normalised) eigenvectors, $\Psi(\alpha)$,  for  these matrices
\begin{equation}
\sum_j H_{ij}\Psi_j(\alpha)=E_{\alpha}\Psi_i(\alpha), \qquad \sum_j \Psi_j(\alpha)\Psi_j(\beta)^*=\delta_{\alpha \beta}\ .
\end{equation}
Then we computed numerically  the distributions of eigenvector components (histograms) for different matrix dimensions and many realisations of random matrices. 

\begin{figure}
\begin{center}
\includegraphics[width=.9\linewidth]{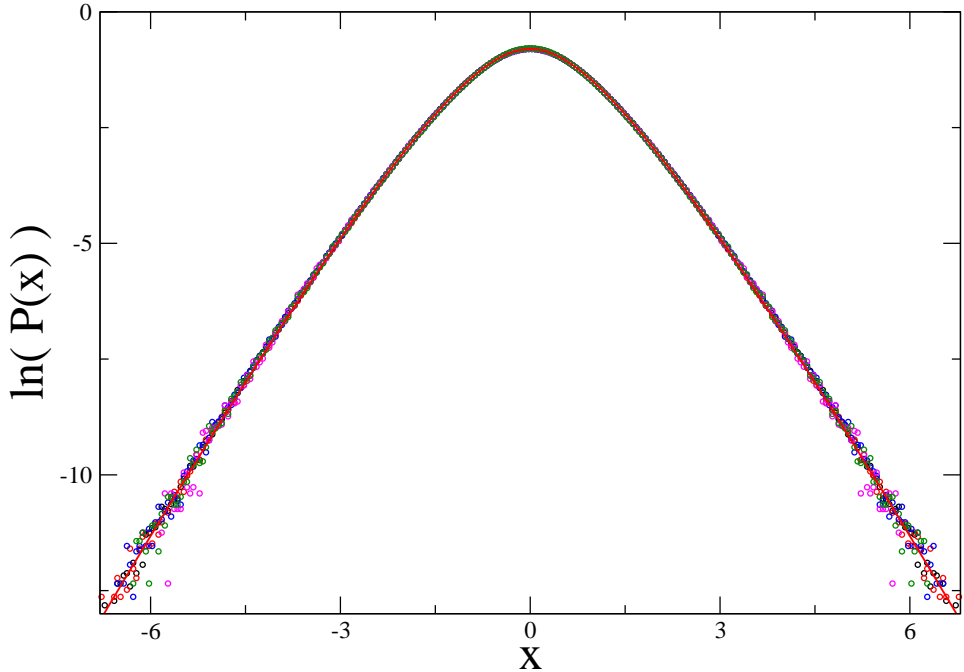}
\end{center}
\caption{Logarithm of the eigenfunction distribution \eqref{norm} for PLBM with $s=0.7$ and $\epsilon=1$ with energies in one half of the spectrum around $E=0$. Data for five different matrix dimensions are superimposed. 
Black circles: $N=8192$, red circles: $N=4096$, blue circles: $N=2048$, green circles: $N=1024$, and magenta circles: $N=512$. The solid red line is the logarithm of the generalised hyperbolic distribution  \eqref{ghd}  with $\alpha=2.6154$, $\lambda=3.3615$, $\delta=0.2903$.} 
\label{eps_1}
\end{figure}

\begin{figure}
\begin{center}
\includegraphics[width=.9\linewidth]{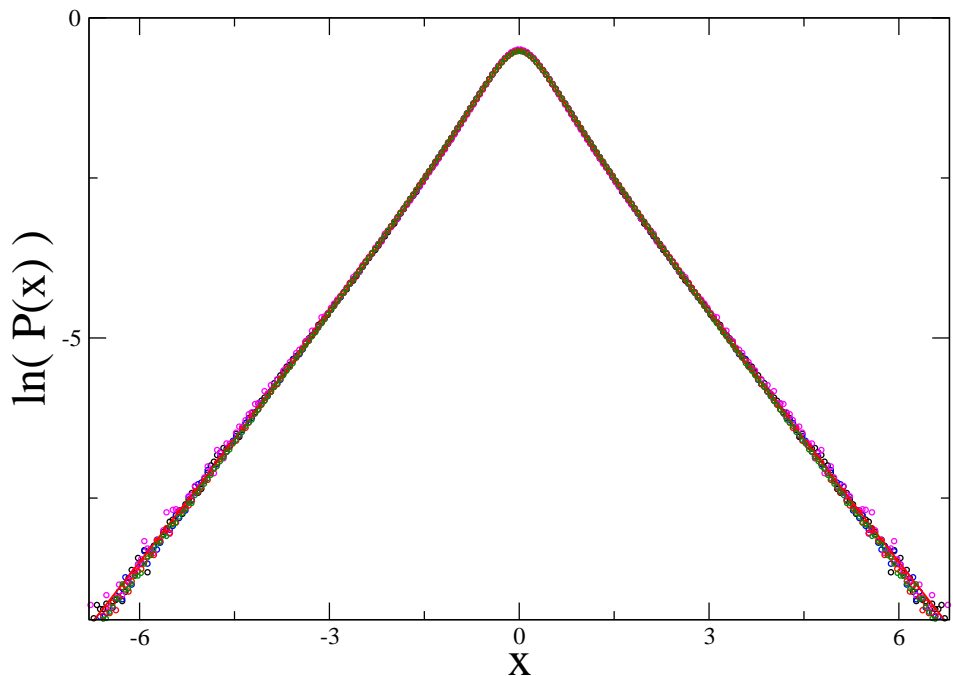}
\end{center}
\caption{The same as in Fig.~\ref{eps_1} but for UMM.  Solid red line: the logarithm of the GHD \eqref{ghd} with $\alpha=1.1673$, $\lambda=0.3880$, $\delta=0.4409$.} 
\label{ln_hierarchical}
\end{figure}

\begin{figure}
\begin{center}
\includegraphics[width=.9\linewidth]{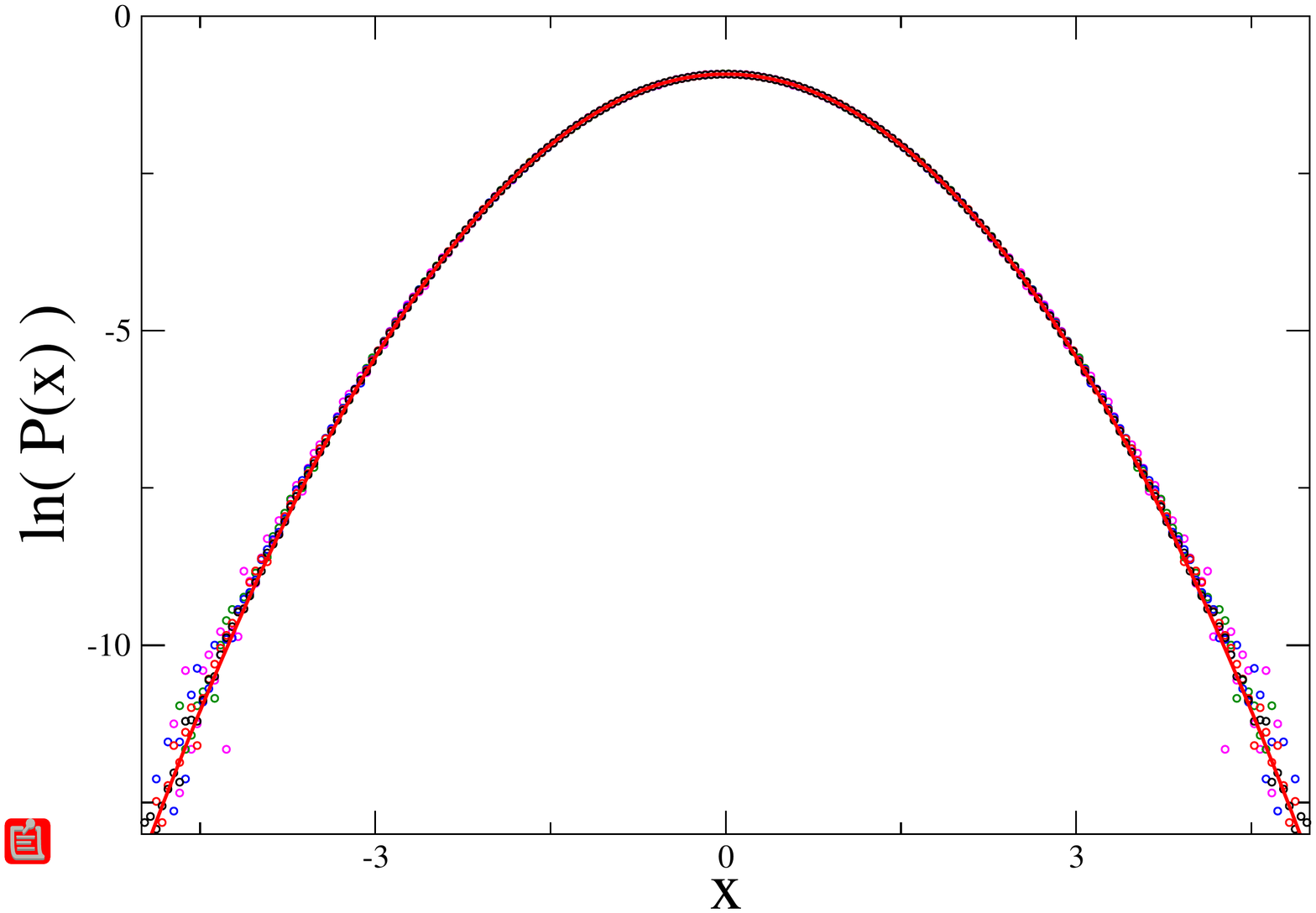}
\end{center}
\caption{The same as in Fig.~\ref{eps_1} but for $s=0.3$.  Solid red line: the logarithm of the  PTD \eqref{pt}. } 
\label{ln_pt}
\end{figure}
Our main results are as follows. 

First, we did not find any indication of  the existence of new phases in these models. Instead, our numerical calculations are fully consistent with completely extended states within the intermediate region \eqref{interval}.  It manifests itself in the fact that  the distribution of the quantity
\begin{equation}
x=\sqrt{N} \Psi_j
\label{norm}
\end{equation} 
becomes quickly independent of the matrix dimension $N$.  As an illustration of this convergence we present in Figs.~\ref{eps_1} and \ref{ln_hierarchical} numerically calculated distributions for PLBM and UMM with $s=0.7$ and $\epsilon=1$ for different values of $N$, $N=2^n$, $n=9-13$. We checked that the distributions do not depend on the component $j$ and we averaged over a few components. The data collapse very quickly to one $N$-independent curve. We also checked that for other values of the parameters $\epsilon$ and $s<1$ the convergence is similar. For comparison  we present in Fig.~\ref{ln_pt} the numerical results for the PLBM but with $s=0.3$ and $\epsilon=1$. For this value of $s$  the model asymptotically corresponds to the GOE and in particular the eigenvector distribution has to coincide with the PTD \eqref{pt}. The figure clearly shows that this is indeed the case.

Second, surprisingly we found that in all considered cases and for both models the eigenvector distribution is extremely well approximated by the so-called generalised hyperbolic distribution (GHD)  which had been introduced to describe mass-size distributions of sand particles \cite{ghd} and was later used mostly in economics.  The probability density function for the GHD (we consider only the symmetric distribution) has  the form 
\begin{equation}
P_{\mathrm{GHD}}(x)=\frac{\sqrt{\alpha}}{\sqrt{2\pi} \delta^{\lambda} K_{\lambda}(\alpha \delta)}\, (x^2+\delta^2)^{(\lambda-1/2)/2}\, K_{\lambda-1/2}\big (\alpha\sqrt {x^2+\delta^2}\big )
\label{ghd}
\end{equation}
where $\alpha,\delta$, and $\lambda$ are free parameters and $K_{\lambda}(x)$ is the K-Bessel function.

This distribution can be considered as the variance mixture of the normal distribution with zero mean and the variance $\sigma^2=y$ distributed according to the generalised inverse Gaussian  distribution (GIG)
\begin{equation}
P_{\mathrm{GHD}}(x)=\int_0^\infty P_{\mathrm{GIG}}(y)\, \frac{\mathrm{e}^{-x^2/2y }}{\sqrt{2\pi y}}\, \mathrm{d}y
\label{mixture}
\end{equation}
where the probability density of the GIG distribution is (see e.g. \cite{gig})
\begin{equation}
P_{\mathrm{GIG}}(x)=\frac{\alpha^{\lambda}}{2 \delta^{\lambda} K_{\lambda}(\alpha \delta)}\, x^{\lambda-1} \mathrm{e}^{-\frac{1}{2} (\alpha^2 x+\delta^2 x^{-1} )}\ .
\label{gig}
\end{equation}
These distributions  have a lot of nice properties, they are both infinitely divisible and their moments can be calculated analytically 
\begin{equation}
C_q\equiv \langle x^{2q} \rangle_{\mathrm{GHD}}=C_{\mathrm{GOE}}(q) \langle x^{q} \rangle_{\mathrm{GIG}},\quad C_{\mathrm{GOE}}(q)=\frac{2^q \Gamma(q+\frac{1}{2})}{\sqrt{\pi}},\quad \langle x^{q} \rangle_{\mathrm{GIG}}=\Big (\frac{\delta}{\alpha}\Big )^q \frac{ K_{\lambda+q}(\alpha\delta)}{K_{\lambda}(\alpha\delta)}\ .
\label{moments}
\end{equation}
Using the properties of the Bessel functions one can check that in the limits  $\lambda\to\pm \infty$ the normalised GHD \eqref{ghd} tends to the PTD \eqref{pt}.

\begin{figure}
\begin{center}
\includegraphics[width=.9\linewidth]{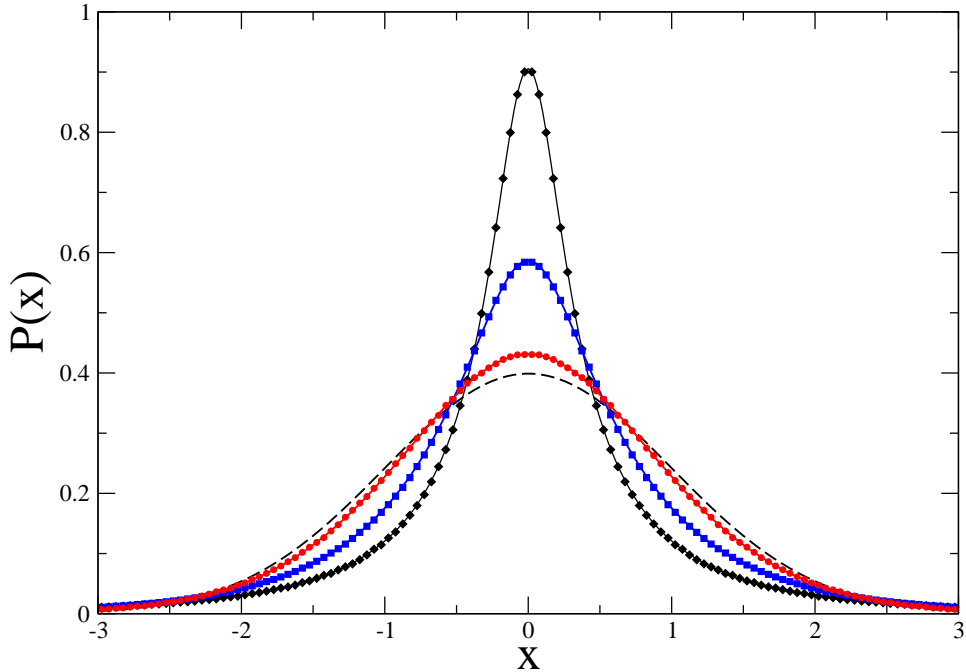}
\end{center}
\caption{Eigenfunction distribution for PLBM with $s=0.7$ and $\epsilon=0.3$  (black circles), $\epsilon=0.5$ (blue squares), and $\epsilon=1.5$ (red diamond) with energies in a half of the spectrum around $E=0$.  Solid  lines with the same colour are fits by the GHD \eqref{ghd} with parameters  $\alpha=0.6506$, $\lambda=-0.1067$, $\delta=0.2805$ (black solid line, for $\epsilon=0.3$), $\alpha=1.2754$, $\lambda=0.5862$, $\delta=0.3945$ (blue solid line, for $\epsilon=0.5$), and $\alpha=2.9341$, $\lambda=3.6392$, $\delta=1.0377$ (red solid line, for $\epsilon=1.5$ ) respectively. The dashed solid line is the PTD \eqref{pt}.} 
\label{fits_different_eps}
\end{figure}

\begin{figure}
\begin{center}
\includegraphics[width=.9\linewidth]{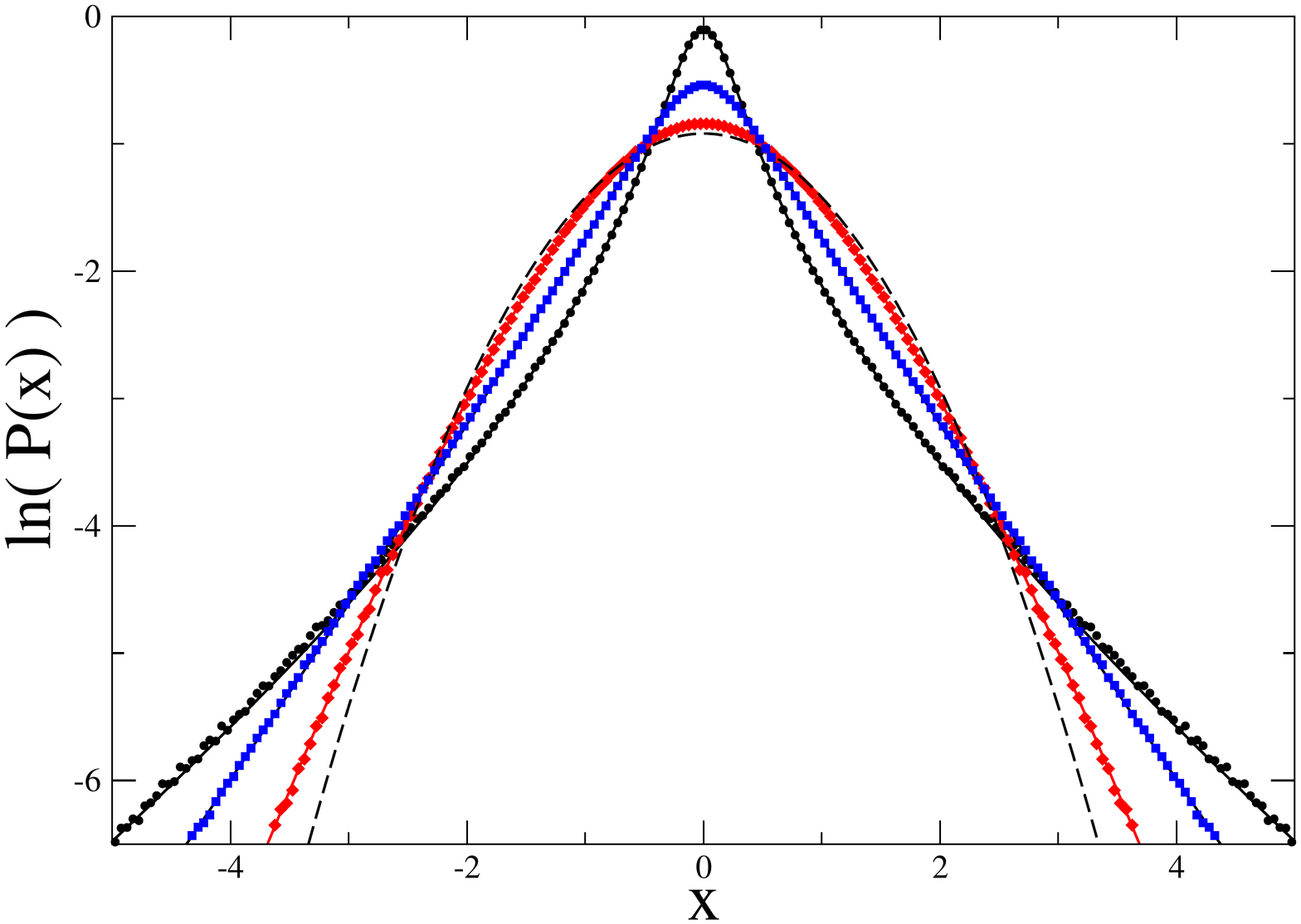}
\end{center}
\caption{The same as in Fig.~\ref{fits_different_eps} but in the logarithmic scale} 
\label{ln_fits}
\end{figure}

\begin{figure}
\begin{center}
\includegraphics[width=.9\linewidth]{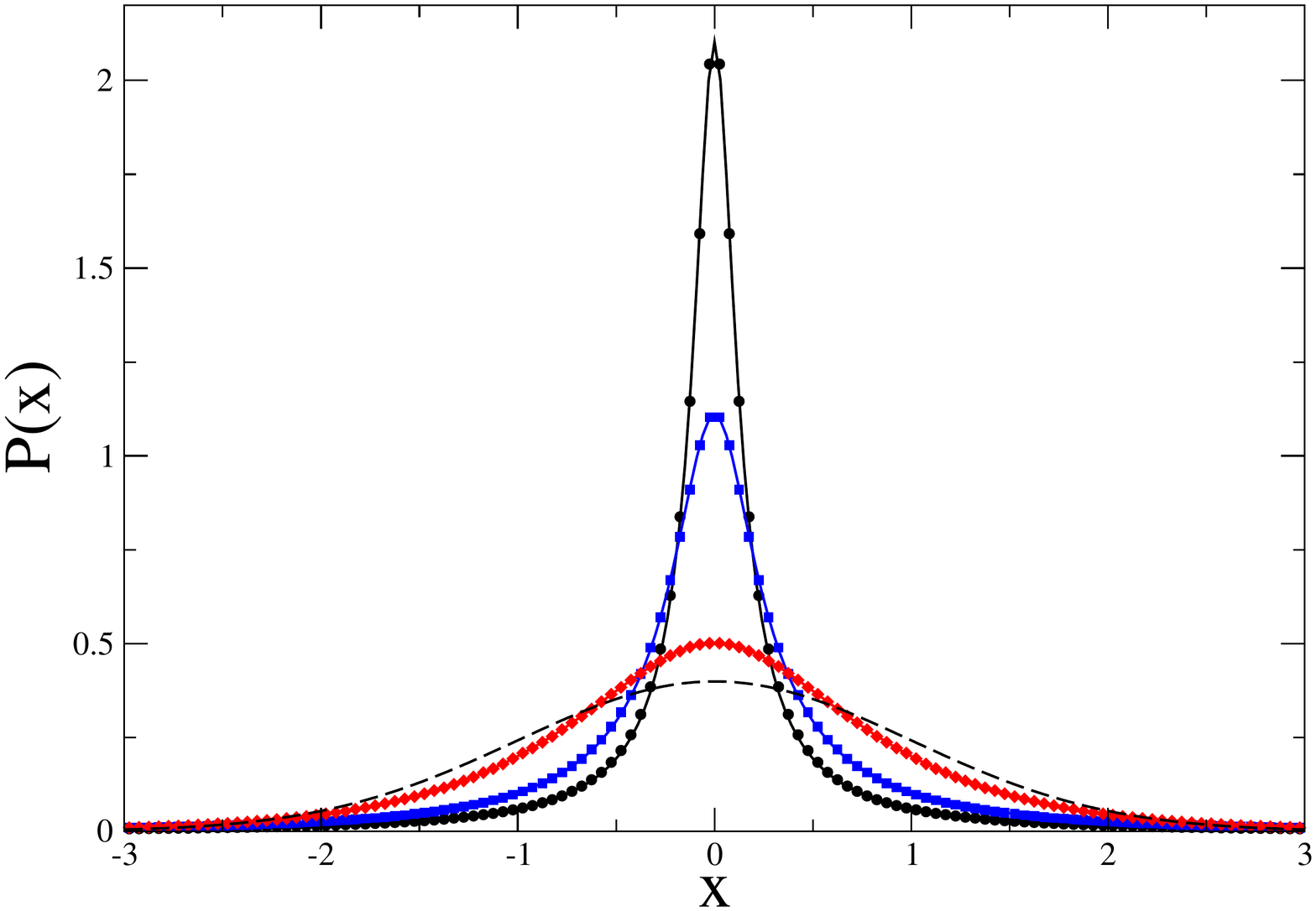}
\end{center}
\caption{The eigenvector distribution for the UMM with $s=0.7$,  $N=2048$ and $1000$ realisations. Black circles: $\epsilon=0.3$, 
blue squares: $\epsilon=0.5$, red diamonds:  $\epsilon=1.5$. Solid lines of the same colour are the GHD fit with parameters: $\alpha=0.2959$, $\lambda=-0.2989$, $\delta=0.1188$ for $\epsilon=0.3$; $\alpha=0.5257$, $\lambda=-0.1857$, $\delta=0.2262$ for $\epsilon=0.5$;
 $\alpha=1.6812$, $\lambda=1.0960$, $\delta=0.5811$ for $\epsilon=1.5$. Dashed line: the PTD \eqref{pt}.} 
\label{bulk_hierarchical}
\end{figure}

\begin{figure}
\begin{center}
\includegraphics[width=.9\linewidth]{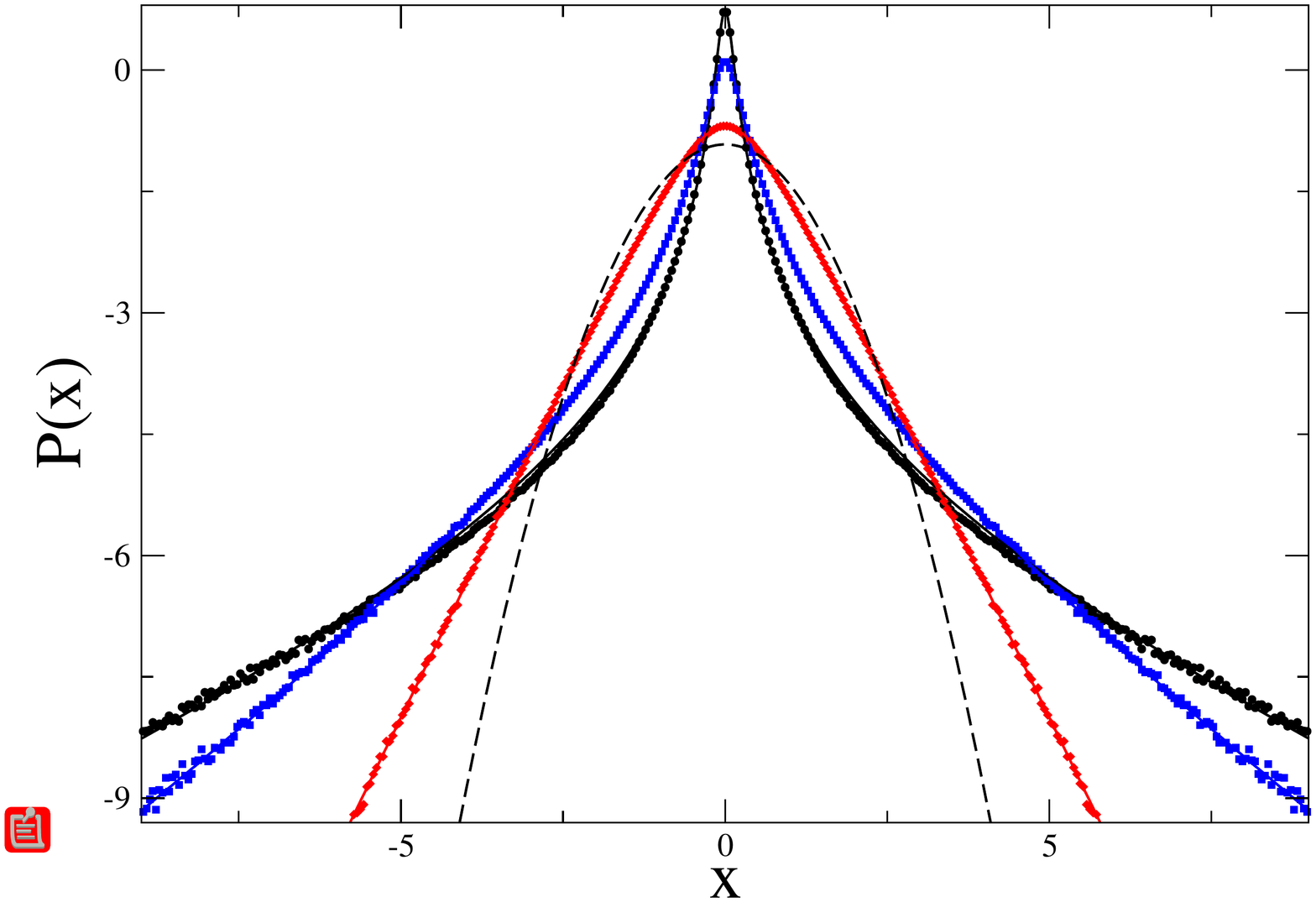}
\end{center}
\caption{The same as in Fig.~\ref{bulk_hierarchical} but in the logarithmic scale}
\label{tail_hierarchical}
\end{figure}

Our numerical procedure was the following. First we calculated numerically eigenvalues and eigenvectors of our matrices for fixed $N$ and different realisations of the random matrix elements with variances as described above. In each run we collected a few  eigenvector components (say the first one) with  eigenvalues in a fixed part of the full spectrum. In the examples below we chose components in half of the spectrum symmetrically around zero energy. Performing a sufficiently large number of realisations (in figures below 1000 realisations were performed) we constructed the normalised histogram of  the chosen eigenvector component multiplied by $\sqrt{N}$ as in \eqref{norm}. A few resulting curves are presented in normal and logarithmic scales for PLBM in Figs.~\ref{eps_1}, \ref{fits_different_eps}, \ref{ln_fits} and for UMM in Figs.~\ref{ln_hierarchical}, \ref{bulk_hierarchical}, and \ref{tail_hierarchical}.

 The PLBM and UMM depend on two parameters, $s$ which determines the decrease of the variance and $\epsilon$ which fixes the pre-factor of the variance. The symmetric GHD depends on 3 parameters, $\alpha, \lambda,\delta$. As by construction eigenvectors are normalised we impose that the variance  of the GHD is equal to 1. From \eqref{moments} it means that 
\begin{equation}
\alpha=\sqrt{\frac{\xi K_{\lambda+1}(\xi)}{K_{\lambda}(\xi)}},\quad \delta=\frac{\xi}{\alpha}, \quad \xi=\alpha \delta\, .
\label{ad}
\end{equation} 
With this substitution the GHD depends on two parameters, $\lambda$ and  $\xi$. We found that for all combinations of the two parameters  $s$ and $\epsilon$ of the PLBM and the UMM  which we investigated, the numerically constructed eigenvector distributions  are   extremely well approximated  by a two parameters fit of the GHD (i.e. by fitting  $\lambda$ and $\xi$).  In Figs.~\ref{eps_1}-\ref{tail_hierarchical} a few examples of such fits are presented.  Notice that with increasing of  $\epsilon$  the resulting distribution tends to the PTD as been argued in \cite{mirlin}. The fits were performed in the normal scale where values of distributions less that 0.01 were removed.  Nevertheless the fits are very good even in the distribution tails as is evident from Figs.~\ref{ln_fits} and \ref{tail_hierarchical}  where the curves are plotted in the logarithmic scale. The agreement with the GHD distribution is remarkable. It is so good that we conjecture that the GHD can be considered as a new universal distribution of extended states on equal footing with the PTD. Further analysis of these models will be discussed  elsewhere. 

The validity of the scaling \eqref{norm} signifies that eigenvectors in both models have the usual 'metallic' fractal dimensions as in the standard RMT
\begin{equation}
\langle \sum_{j=1}^N |\Psi_j|^{2q}\rangle \underset{N\to\infty}{\longrightarrow}N^{-(q-1)}C_q
\end{equation}
 but the pre-factors $C_q$ differ from the values calculated from the PTD and  have to be calculated from the GHD, see \eqref{moments}.

\begin{figure}
\begin{center}
\includegraphics[width=.9\linewidth]{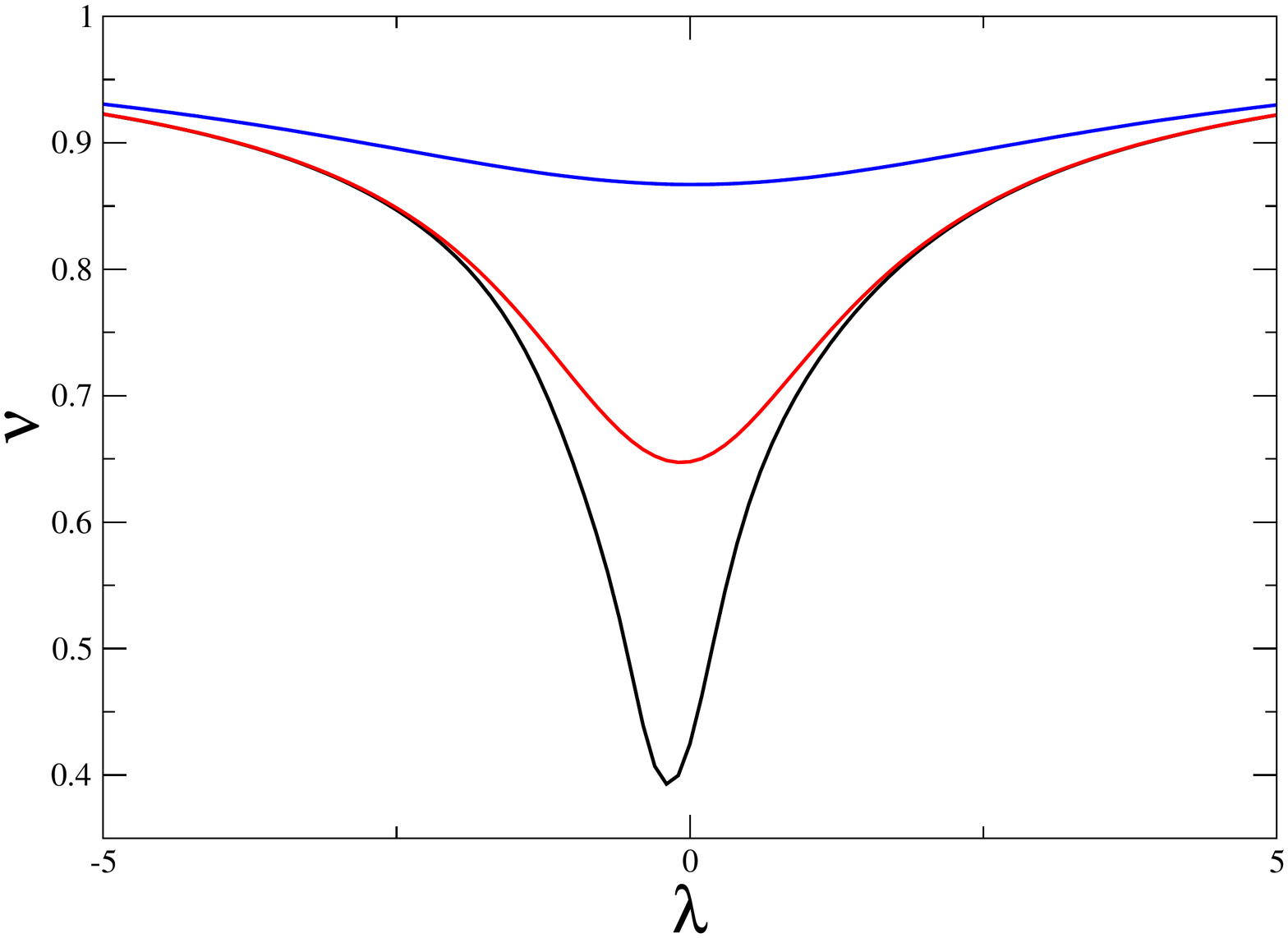}
\end{center}
\caption{The fits of the normalised $\chi^2$-distribution \eqref{chi} with parameter $\nu$ to the normalised GHD \eqref{psi_squared} with fixed $\xi$ and different $\lambda$. From top to bottom, blue, red, and black curves correspond respectively to $\xi=2,\xi=0.2$, and $\xi=0.02$.} 
\label{fig_chi}
\end{figure}

To compare our distribution with experimental results \cite{koehler_1}-\cite{koehler_3} we fitted the normalised GHD \eqref{ghd}, \eqref{ad} with
the normalised  $\chi^2$-distribution dependent on one parameter $\nu$
\begin{equation}
P_{\chi^2}(x,\nu)=\frac{\nu^{\nu/2}\, x^{\nu/2-1} }{2^{\nu/2}\Gamma(\nu/2)}\mathrm{e}^{-\nu\, x/2}, \qquad \langle x \rangle_{\chi^2}=1\ .
\label{chi}
\end{equation} 
As the $\chi^2$-formula corresponds to the distribution   of eigenfunctions squared, $x=\Psi^2$, we transformed \eqref{ghd} to this variable
\begin{equation}
P(\Psi^2=x)=\frac{P_{\mathrm{GHD}}(\sqrt{x})}{\sqrt{x}}. 
\label{psi_squared}
\end{equation}
The normalised GHD depends on 2 parameters $\lambda$ and $\xi=\alpha\delta$. We chose a few values of  $\xi=0.02,\, 0.2,\, 2$ and for different  $\lambda$ found the best non-linear fit of  the normalised $\chi^2$-distribution \eqref{chi}  to the normalised Eq.~\eqref{psi_squared} in the interval $[0.04\ldots 4]$.  The results are plotted at Fig.~\ref{fig_chi}. Though the quality of these fits  is variable,  it follows that the GHD can  be reasonably well approximated by the $\chi^2$-distribution with parameter $\nu$ in the intervals obtained in experimental studies of the neutron widths for different isotopes  ($\nu=0.57\pm 0.16 $ for ${}^{192}\mathrm{Pt}$,  $\nu=0.47\pm 0.19 $ for ${}^{194}\mathrm{Pt}$, $\nu=0.60\pm 0.28 $ for ${}^{196}\mathrm{Pt}$) \cite{koehler_1}-\cite{koehler_3}. We think that our approach permits to explain naturally the observed deviations from the PTD without additional assumptions.

%====================================
\section{Mean local eigenvector variance}

\begin{figure}
\begin{center}
\includegraphics[width=.9\linewidth]{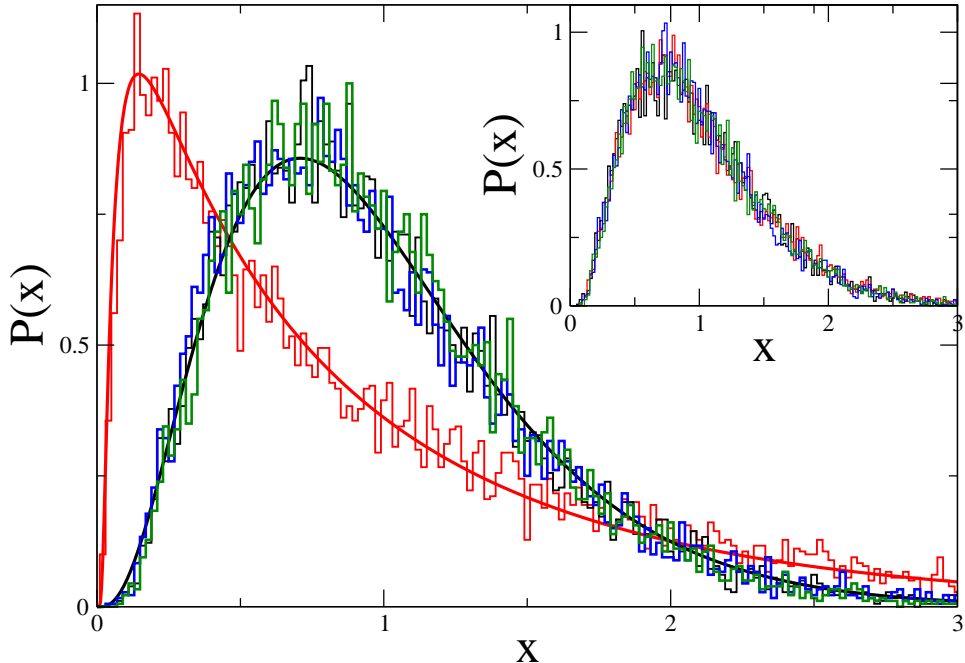}
\end{center}
\caption{ Mean local eigenvector variance \eqref{sigma}  for the PLBM  with $N=4096$ and $1000$ realisations. 
Red staircase line (closest to the ordinate axis):  $s=0.7$, $\epsilon=0.5$, $M_I=100$.
Other staircase lines correspond to $s=0.7$ and $\epsilon=1$ for different $M_I$: blue line $M_I=50$, black line $M_I=100$, and green line $M_I=200$.
Solid red and black  lines  indicate the GIG distribution \eqref{gig} with parameters taken from the fits of eigenvector distributions (see the captions to Figs.~\ref{eps_1} and \ref{fits_different_eps})).
Insert: The mean local eigenvector variance for  $s=0.7$, $\epsilon=1$, $M_I=100$ but for different matrix dimensions. Black line $N=1024$, red line $N=2048$, blue line $N=4096$, and green line $N=8192$.}
\label{variance_PLBM}
\end{figure}

\begin{figure}
\begin{center}
\includegraphics[width=.9\linewidth]{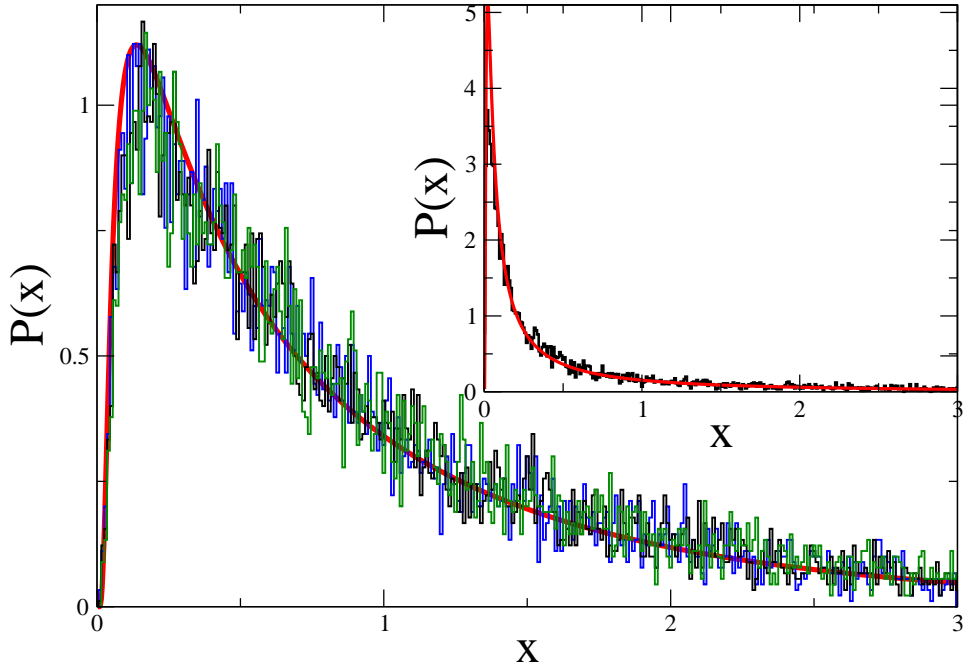}
\end{center}
\caption{Mean local eigenvector variance \eqref{sigma}  for the UMM with $N=4096$,  $s=0.7$, $\epsilon=1$, and $1000$ realisations. Blue staircase line: $M_I=50$, black staircase line: $M_I=100$, and green staircase line: $M_I=200$. Solid red line  indicates the GIG distribution \eqref{gig} with parameters taken from the fits of eigenvector distributions (see the caption to Fig.~\ref{ln_hierarchical}). Insert:  Black staircase line is the mean local eigenvector variance  for UMM for $N=4096$, $s=0.7$,  $M_I=100$  but for $\epsilon=0.5$. Solid red  is the GIG distribution from the fit indicated in Fig.~\ref{bulk_hierarchical})).}
\label{variance_UMM}
\end{figure}

If the observed eigenvector distribution  is indeed the variance mixture as in Eq.~\eqref{mixture}, then the eigenvector variance should be distributed according to  the GIG distribution \eqref{gig} (or at least be close to it). To calculate this quantity the following method was used. 
Choose  a fixed energy  interval $I=[E-\delta E/2, E+\delta E/2]$  where  the width $\delta E$ is such that the interval contains $M_I\gg 1$ eigenvalues but all global quantities (such as the mean level density) remain practically constant. Calculate the mean value of the eigenvector variance for each realisation as follows
\begin{equation}
x=\frac{1}{M_I}\sum_{E_{\alpha}\in I} N  |\Psi_i(E_{\alpha})|^2
\label{sigma}
\end{equation}
where $i$ is an arbitrary eigenvector component. This is a random number and collecting it for different realisations permits to find its distribution numerically. In Figs.~\ref{variance_PLBM}  and \ref{variance_UMM} the results for the PLBM and UMM with $N=4096$ are presented. To diminish the fluctuations the data for 9 different eigenvector components (with indices $i=1,N/8,2N/8,\ldots, N$) were combined for $M_I$ nearest  eigenvalues around the centre of the spectrum ($E=0$). The corresponding GIG distributions with parameters obtained from the fits of eigenvector distributions in Figs.~\ref{eps_1}, \ref{ln_hierarchical}, \ref{fits_different_eps} and \ref{bulk_hierarchical} are also plotted in these figures. Though the fluctuations of the variances are bigger than the ones for the eigenvector distributions  (due to lower statistics), the numerical data for the variances are in a good agreement with the GIG prediction thus confirming the variance mixture origin of the GHD. 

The result suggests that the GHD appears in the considered models after a two-step averaging procedure. First one averages eigenvectors   over a small energy window as in Eq.~\eqref{sigma} with fixed matrix elements and then finds  the distribution of  their variance  for different realisations. The full eigenvector distribution corresponds to the local Porter-Thomas law (the Gaussian) with this random variance. Similar considerations were applied to the Rosenzweig-Porter model   in \cite{sieber}.  

\begin{figure}
\begin{center}
\includegraphics[width=.9\linewidth]{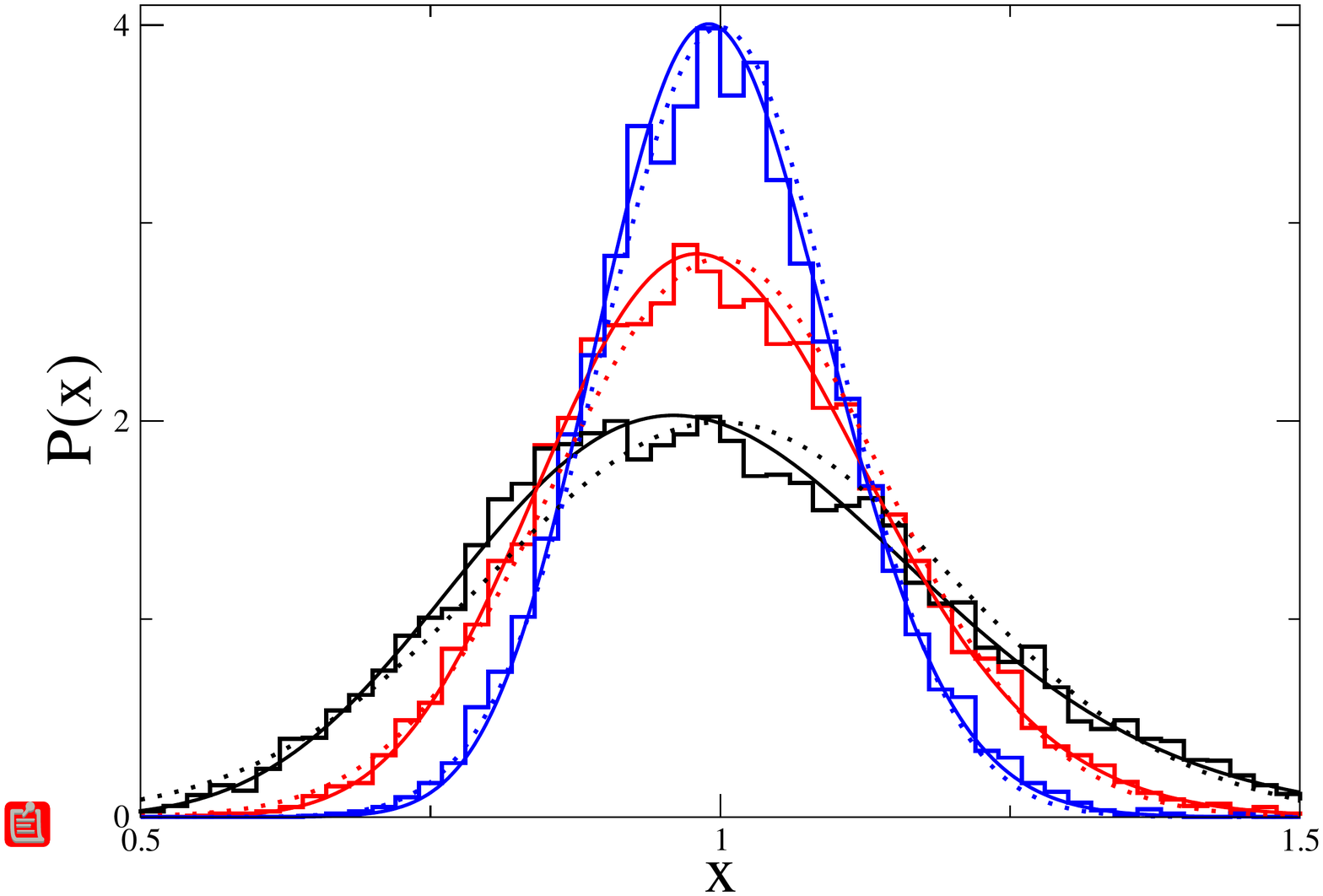}
\end{center}
\caption{Mean local eigenvector variance \eqref{sigma}  for the PLBM with $s=0.3$ with $\epsilon=1$ and different number of eigenvalues $M_I$. 
From top to bottom at $x=1$: blue staircase line: $M_I=200$, red staircase line: $M_I=100$, black staircase line: $M_I=50$. Solid lines of the same colour are normalised $\chi^2$ distributions with $\nu=M_I$ degrees of freedom \eqref{chi}. Dotted lines of the same colour  are the asymptotic  Gaussian approximations to these distributions \eqref{asymptotic_chi}.}
\label{variance_goe}
\end{figure}

It is important to stress the difference between the results for the intermediate interval  $\frac{1}{2}<s<1$ and the (usual) GOE case $s<\frac{1}{2}$.  For the latter case  the eigenvector distribution  should be described by  the energy-independent PTD \eqref{pt}. As in the sum \eqref{sigma} all terms are iid Gaussian random  variables, the variance  \eqref{sigma} has to be distributed as the normalised $\chi^2$ distribution \eqref{chi} with parameter $\nu$ equal $M_I$. For large $M_I$ one can use the usual asymptotic expression for the sum \eqref{sigma} which follows from the central limit theorem
\begin{equation}
P(x)_{\mathrm{GOE}}\underset{M_I\to\infty}{\longrightarrow} \sqrt{\frac{M_I}{4 \pi}}\mathrm{e}^{-M_I(x-1)^2/4} \, .
\label{asymptotic_chi}
\end{equation}
In Fig.~\ref{variance_goe} we present the mean variance for the PLBM with $s=0.3$ (and $\epsilon=1$). It is seen that the above prediction is very well confirmed by numerics. 

 The situation for the intermediate regime $\frac{1}{2}<s<1$ is completely different. The data with different $M_I$ show no tendency to shrink with increasing of $M_I$. To illustrate it, in Figs.~\ref{variance_PLBM} and \ref{variance_UMM} the mean local eigenvector variance are shown for  different values of $M_I=50,\ 100,\ 200$ (the same as in Fig.~\ref{variance_goe}).  It is clearly seen that  the curves are superimposed and no noticeable changes are observed. For completeness, in the insert of Fig.~\ref{variance_PLBM} the data with $M_I=100$ but for different $N$ are plotted.  As expected,  the data demonstrate  no changing  with $N$. 
 
 The numerically observed independence of the local variance on the number of successive eigenvalues included in its calculation implies
 that eigenvectors with close  eigenvalues are not independent. Otherwise the central limit theorem will force the distribution to be asymptotically Gaussian similar to  Eq.~\eqref{asymptotic_chi} which seems not to be the case. Another manifestation of the same phenomenon is that  the Green function in the  intermediate regime \eqref{interval} (at least its  imaginary part) is not  a self-averaged quantity as it is in all models considered so far. Further investigation of  these questions will be discussed elsewhere.  
 
%====================================
\section{Conclusions}

Power-law banded and ultrametric matrices are typical representatives of random matrix ensembles with varying strength of interaction. Contrary to the usual random matrix ensembles these ensembles are constructed in such a way that the interaction between two sites decreases  as a certain power, $s$, of the distance between the sites. If the interaction  decays quickly ($s>1$) off-diagonal terms play a minor role, eigenfunctions are localised, and eigenvalues behave as iid random variables. In the opposite limit ($s<\frac{1}{2}$) eigenfunctions  are fully delocalised and the spectral statistics is close to the standard random matrix statistics.  In both models there exits an interval of parameters where eigenfunctions are supposed to be delocalised but their properties remain elusive because known analytical tools are inapplicable (or at least not developed). For PLBM and UMM this intermediate regime corresponds to $\frac{1}{2}<s<1$.   

We investigated numerically the eigenvector distribution for PLBM and UMM in this interval for various combinations of model parameters. No anomalous scaling was observed. Our findings indicate that after rescaling  by $\sqrt{N}$ eigenvector distributions  become $N$-independent functions which implies that fractal dimensions are the same as for the usual RMT. 

 Our main result is the observation that in all considered cases the eigenvector distributions   can be  extremely accurately fitted  by the generalised hyperbolic distribution  which  differs considerably from the Porter-Thomas distribution which is the standard result in RMT. This universality is intriguing as the non-existence of analytical approaches  in the intermediate regime suggests that  all quantities may depend on details of the models and not be universal.   
 
 A direct application of this result is the possibility that the observed experimental  deviations of recent experimental  data of neutron  widths from the PTD could  be naturally  explained by the physically reasonable  assumption  that the corresponding RM model  for neutron  widths is not of usual GOE type  but belongs to the PLBM. 
 
The investigation of the PLBM and UMM in the intermediate regime seems to be overlooked but is of importance as they constitute a new class of random matrices potentially important for different applications.

%====================================

\begin{acknowledgments}
The authors are indebted to Y. Fyodorov, J. Keating, J. Marklof, and Y. Tourigny for many useful discussions. 
One of the authors (EB) is grateful to  the Institute of Advanced Studies at the University of Bristol for financial support in form of a Benjamin Meaker Visiting Professorship and the School of Mathematics for hospitality during the visit where this paper was written. 
 \end{acknowledgments}

%====================================

\end{document}